\documentclass[
 reprint,
superscriptaddress,
 amsmath,amssymb,
 aps,
]{revtex4-2}
\usepackage{mathtools}
\usepackage{graphicx}% Include figure files
\usepackage{dcolumn}% Align table columns on decimal point
\usepackage{bm}% bold math
\usepackage{siunitx}
\usepackage[utf8]{inputenc}
\usepackage[T1]{fontenc}

\makeatletter
\renewcommand*{\fnum@figure}{{\normalfont\bfseries \figurename~\thefigure}}
\renewcommand*{\@caption@fignum@sep}{\textbf{\usepackage{.} }}
\makeatother
\usepackage{amsmath}
\usepackage {algpseudocode}
\usepackage{lipsum}% for dummy text
\makeatletter
\renewcommand*{\fnum@figure}{{\normalfont\bfseries \figurename~\thefigure}}
\renewcommand*{\@caption@fignum@sep}{\textbf{. }}
\makeatother
\renewcommand{\thefigure}{\arabic{figure}}

\begin{document}

\preprint{APS/123-QED}
\raggedbottom
\title{Computational Scaling in Inverse Photonic Design Through Factorization Caching}% Force line breaks with \\

\author{Ahmet Onur Dasdemir}
\email{adasdemir16@ku.edu.tr}
\affiliation{%
 Department of Electrical and Electronics Engineering, Koç University\\
Sariyer, Istanbul, 34450, Turkey
}%

\author{Victor Minden}
\affiliation{%
Path AI, 1325 Boylston St Suite 10000, Boston, MA 02215, USA
}%
\author{Emir Salih Magden}
\email{Corresponding author: esmagden@ku.edu.tr}
\affiliation{%
 Department of Electrical and Electronics Engineering, Koç University\\
Sariyer, Istanbul, 34450, Turkey
}%

%%\date{\today}% It is always \today, today,
             %  but any date may be explicitly specified

\begin{abstract}
Inverse design coupled with adjoint optimization is a powerful method to design on-chip nanophotonic devices with multi-wavelength and multi-mode optical functionalities. Although only two simulations are required in each iteration of this optimization process, these simulations still make up the vast majority of the necessary computations, and render the design of complex devices with large footprints computationally infeasible. Here, we introduce a multi-faceted factorization caching approach to drastically simplify the underlying computations in finite-difference frequency-domain (FDFD) simulations, and significantly reduce the time required for device optimization. Specifically, we cache the symbolic and numerical factorizations for the solution of the corresponding system of linear equations in discretized FDFD simulations, and re-use them throughout the entire device design process. As proof-of-concept demonstrations of the resulting computational advantage, we present simulation speedups reaching as high as  $9.2\times$ in the design of broadband wavelength and mode multiplexers compared to conventional FDFD methods. We also show that factorization caching scales well over a broad range of footprints independent of the device geometry, from as small as \SI{16}{\micro\metre ^2} to over \SI{7000}{\micro\metre ^2}. Our results present significant enhancements in the computational efficiency of inverse photonic design, and can greatly accelerate the use of machine-optimized devices in future photonic systems.

\end{abstract}

%\keywords{Suggested keywords}%Use showkeys class option if keyword
                              %display desired
\maketitle

%\tableofcontents

\section{Introduction}

The increasing number and complexity of optical operations required in communications, sensing, and computing applications continue to drive the development of compact and efficient integrated photonic components \cite{yang2022multi, luan2018silicon, vandoorne2014experimental}. Demonstration of these high-performance on-chip devices with multi-wavelength and multi-mode capabilities has significant implications for the throughput and scalability of next-generation optical systems \cite{piggott2015inverse, lu2012objective, magden2018transmissive, singh2020silicon}. As the demand for such high-performance, multi-function, and compact on-chip devices continues to grow, traditional design approaches that rely on physical intuition and trial-and-error have become insufficient in meeting these stringent requirements. Consequently, automated design algorithms have recently attracted significant attention and are rapidly proving to be indispensable tools for photonic design. To this end, inverse design techniques have emerged as a powerful approach to design photonic devices with unprecedented performance and functionality by exploring a large set of design variables \cite{piggott2015inverse, molesky2018inverse, hughes2018adjoint, lalau2013adjoint, lu2012objective}. Previous research on inverse design has employed two main techniques to achieve optimal device structures: shape optimization \cite{lalau2013adjoint, michaels2018leveraging, michaels2019hierarchical, piggott2017fabrication}, which involves modifying geometrical boundaries within a specified design region, and topology optimization \cite{piggott2015inverse, hughes2018adjoint, elesin2014time, hammond2022high, zhang2021topological}, which updates the device in a pixel-by-pixel manner. In both of these methods, the device structure is modified iteratively by optimizing an objective function that is calculated from the finite-difference frequency-domain (FDFD) solutions of Maxwell’s equations. The gradient of this objective with respect to the design variables is obtained through an adjoint system, and then used to gradually modify the photonic device through gradient-based optimization methods \cite{hughes2018adjoint,lalau2013adjoint,michaels2018leveraging,piggott2015inverse}.

Although only two simulations are required for each iteration of this optimization process, these "forward" and "adjoint" physical simulations in optical inverse design problems are known to be computationally expensive, making up the vast majority of the total optimization time \cite{zhao2019accelerating}. To address this issue, prior work included accelerating individual simulations by reducing the size of the system matrix in FDFD via Schur complement decomposition \cite{zhao2019accelerating}. This method works well when the optimizable design region constitutes a relatively small portion of the simulation domain, but is less effective for devices with larger footprints. Moreover, the overhead required for an initial system solution can be computationally prohibitive, which renders this method only applicable to a limited subset of photonic design problems. On the other hand, deep learning-enabled approaches replacing Maxwell’s solvers have also been proposed \cite{ma2021deep, chen2022high, liu2018training}. Even though such learned representations can be more computationally efficient than traditional linear system solutions, these methods can suffer from the lack of physical accuracy due to the loss of generalization when designing devices significantly different from those used in their training datasets \cite{neyshabur2017exploring}. Hence, addressing the computational requirements for accurate device simulations remains a critical challenge, particularly as the total optimization times increase substantially with larger device footprints and greater number of iterations needed for convergence. These computational requirements prohibit the design of large-scale and multi-purpose devices for next-generation photonic systems with complex functionalities. Therefore, significant progress in computational capabilities is necessary in order to enable and streamline physically accurate and data-independent procedures for the design of large-scale integrated optical devices with novel, previously-elusive, and multi-purpose functionalities.

Here we demonstrate a set of factorization caching approaches that drastically improve the computational efficiency of FDFD simulations in photonic inverse design. We utilize the underlying patterns in the corresponding set of discretized Maxwell’s equations, cache factorizations of the resulting system matrices, and re-use them throughout the design process. With this method, we demonstrate a substantial speedup in FDFD simulations, enabling faster and more efficient optimizations for large-scale and multi-purpose photonic device design. As proof-of-concept devices, we design a wavelength duplexer and a mode multiplexer using our caching methods, and demonstrate enhancements of up to 9.2-fold in simulation speeds over conventional FDFD techniques. Our approach demonstrates how computational enhancements can be leveraged to accelerate physical simulations and enable exciting new possibilities for large-scale on-chip photonic design.

\section{Computational Motivation for Factorization Caching}

We first begin with a brief overview of adjoint optimization and specifically its use in nanophotonics. The goal of inverse photonic design is to create an ideal device geometry by optimizing an objective function $F(\mathbf{E})$ using the electric field $\mathbf{E}$ computed from Maxwell's equation
\begin{equation}
    \left( - \omega^2\varepsilon_0\varepsilon_r(\mathbf{r}) +{{\mu_{0}}^{-1}}\nabla\times \nabla\times \right) \mathbf{E(r)} = -i\omega \mathbf{J(r)}
\label{maxwell_eq}
\end{equation}
where $\varepsilon_r(\mathbf{r})$ is the relative permittivity distribution representing the design variables $\theta_i$, and $\mathbf{J(r)}$ is the input source. FDFD simulations solve a discretized version of Eq.~\ref{maxwell_eq} as a linear system of equations in the form of 
\begin{equation}
A \mathbf{E} = -i\omega \mathbf{J}
\label{Ax=b}
\end{equation}
where $A$ is the system matrix including terms with permittivity distribution and spatial derivatives. Once the linear system in Eq.~\ref{Ax=b} is solved and the design performance calculated by the objective function $F(\mathbf{E})$, the gradient of this objective with respect to the design variables $\theta_i$ is computed using the adjoint system \cite{hughes2018adjoint,lalau2013adjoint,michaels2018leveraging}
\begin{equation}
    \frac{\partial F}{\partial \theta_i} = -2\mathrm{Re}\{\mathbf{E_{adj}}^{T} \frac{\partial A}{\partial \theta_i}\mathbf{E} \}
\end{equation}
and $\mathbf{E_{adj}}$ is obtained by solving the linear set of equations given by
\begin{equation}
A^T \mathbf{E_{adj}} = \left(\frac{\partial F}{\partial \mathbf{E}}\right) ^T
\label{adjoint_eq}
\end{equation} 
For solutions of Eq.~\ref{Ax=b} and Eq.~\ref{adjoint_eq}, most direct solvers use LU factorization to decompose $A$ into its lower and upper triangular factors as $A=LU$ such that the solutions can be formulated as 
\begin{equation}
 \mathbf{E} = -i\omega U^{-1}L^{-1} \mathbf{J}
\label{back_substitution}
\end{equation}
and
\begin{equation}
 \mathbf{E_{adj}} = (L^T)^{-1}  (U^T)^{-1} \left(\frac{\partial F}{\partial \mathbf{E}}\right) ^T
\label{adjoint_solution}
\end{equation}

Considering the structure of the individual terms in Eq.~\ref{maxwell_eq}, we note that $A$ is a sparse matrix, and its LU factorization introduces fill-ins (new nonzero entries) in the $L$ and $U$ factors. Symbolic factorization is the process in which the locations of these fill-ins and their analytical expressions are determined, according to the sparsity pattern of $A$. Following symbolic factorization, the numerical values of these entries are calculated via a subsequent numerical factorization \cite{davis2016survey}. Once these factorizations have been completed, the final solution to the sparse linear system is obtained through back substitution as shown in Eq.~\ref{back_substitution}. 

\begin{figure}[t]
\centering\includegraphics[width=\linewidth]{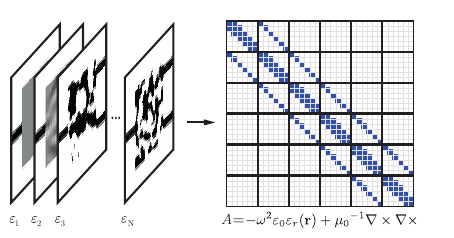}
\caption{Sparsity pattern of the system matrix $A$ is independent of the spatial distribution of  materials, and remains the same throughout the entire design procedure.}
\label{fig:figure1}
\end{figure}
Our factorization caching approach is motivated by the structure of and the updates on the system matrix $A$, and its corresponding $L$ and $U$ factors. First of all, during a single iteration of optimization, numerical factors of  $A^T$, namely  $U^T$ and  $L^T$ are required to solve the adjoint system in Eq.~\ref{adjoint_eq}. Fortunately, once $L$ and $U$ are numerically calculated, $L^{T}$ and $U^{T}$ are already known, making the solution of Eq.~\ref{adjoint_solution} trivial through back substitution. As such, no separate numerical factorizations are necessary for the adjoint system solution in the same iteration. This is extremely advantageous since solving a system through back substitution can be over 10 times faster than re-factoring $A^T$ from scratch \cite{zhao2022efficient}. 
Secondly, as depicted in Fig.~\ref{fig:figure1}, the sparsity pattern of $A$ is dictated by the spatial derivative terms in Eq.~\ref{maxwell_eq}. Consequently, as the device geometry is iteratively modified during optimization, while $A$ is updated numerically, its sparsity pattern remains the same. As such, even though consecutive system solutions need to be performed using numerically different $A$ matrices, each one of these solutions can utilize the same symbolic factorization of $A$. This symbolic factorization needs to be computed only once at the beginning, and can be reused throughout the rest of the optimization process. The same symbolic factorization can even be used regardless of wavelength, which is especially beneficial for designing multi-wavelength photonic devices. As wavelength dependence in permittivity $\varepsilon_r(\mathbf{r})$ and in $\omega$ only alter $A$ numerically, this stored symbolic factorization can be repeatedly reused for FDFD simulations at different wavelengths. 
Finally, we note that these advantages scale well to devices that process multiple sets of inputs such as spatial mode multiplexers. For these cases, systems with different right hand sides (i.e. different $\mathbf{J}$ inputs) can be solved using the existing $L$ and $U$ factors, as the device geometry and the corresponding system matrix $A$ remain entirely the same across different inputs during a single iteration. The combination of these three separate factorization caching approaches can enable significant computational scaling in the design of large-scale and multi-purpose integrated photonic devices. In the following sections, we illustrate these resulting computational advantages for two example classes of nanophotonic devices with wavelength-dependent and spatial mode-dependent functionalities.

\section{Multi-Wavelength Photonic Duplexer Demonstration with Factorization Caching}

\begin{figure}[t]
\centering\includegraphics[width=\linewidth]{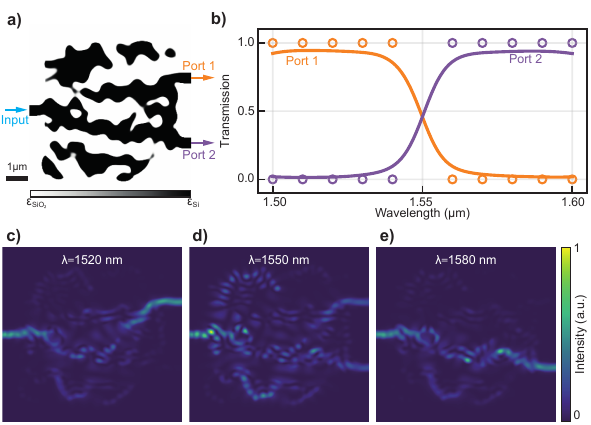}
\caption{\textbf{(a)} Wavelength duplexer geometry. \textbf{(b)} Transmission targets (shown with colored circles) and the transmission at output ports. \textbf{(c-e)} Electric field intensity distribution of wavelength duplexer at \SI{1520}{\nano\metre}, \SI{1550}{\nano\metre} and \SI{1580}{\nano\metre}.}
\label{fig:figure2}
\end{figure}

Our factorization caching approach is first applied in the design of a broadband silicon photonic wavelength duplexer with a \SI{7}{\micro\metre}$\times$\SI{7}{\micro\metre} device footprint, separating the input light into short-pass (1.50-\SI{1.55}{\micro\metre}) and long-pass (1.55-\SI{1.60}{\micro\metre}) output ports. The objective function $F(\mathbf{E})$ is defined as the difference between the calculated and desired transmissions at 10 target wavelengths, and iteratively minimized for broadband operation. Fig.~\ref{fig:figure2}(a) shows the final device design, for which a smoothing filter with a \SI{200}{\nano\metre} radius and a sigmoid-like projection were used to binarize the device geometry to $\varepsilon_{\mathrm{Si}
}$ and $\varepsilon_{\mathrm{SiO_2}}$ permittivities \cite{dasdemir2020multi}. The transmission targets and the final device’s FDFD results at both outputs are plotted in Fig.~\ref{fig:figure2}(b), where the device is shown to perform the spectral duplexing operation as specified with insertion loss as low as 0.26dB (94.1\% transmission). For optimization, the minimization of $F(\mathbf{E})$ is performed through an L-BFGS-B optimizer \cite{zhu1997algorithm} where forward and adjoint FDFD simulations are run at all target wavelengths in each iteration. The resulting electric field profiles at \SI{1.52}{\micro\metre}, \SI{1.55}{\micro\metre} and \SI{1.58}{\micro\metre} are shown in Fig.~\ref{fig:figure2}(c) through Fig.~\ref{fig:figure2}(e).

\begin{figure}[t]
\centering\includegraphics[width=\linewidth]{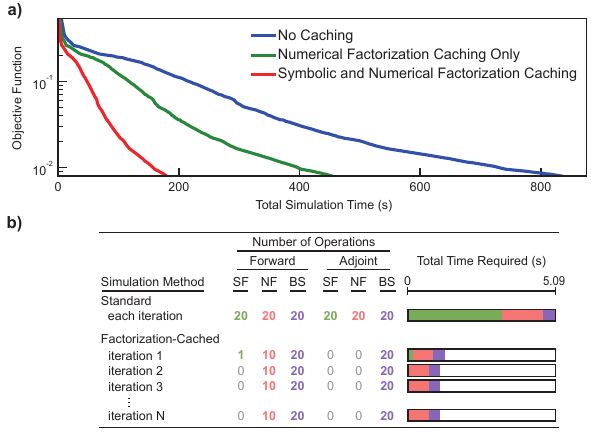}
\caption{\textbf{(a)} Objective function throughout the optimization of a wavelength duplexer as a function of total simulation time with different factorization caching approaches. \textbf{(b)} Numbers of symbolic factorization (SF), numerical factorization (NF) and back substitution (BS) operations in each iteration for no caching and our factorization caching method. Bar plots visualize the total simulation and the corresponding SF, NF and BS runtimes per iteration.}
\label{fig:figure3}
\end{figure}

In order to demonstrate the utility and the computational advantage of our factorization caching approach, the device design is repeated three times, where we progressively apply numerical and symbolic factorization caching. In Fig.~\ref{fig:figure3}(a), we plot the objective function with respect to the total simulation time spent through 164 iterations, for a target objective of $8 \times 10\textsuperscript{-3}$. All simulations were performed on a PC with a 2.4 GHz Intel Xeon Gold processor with 8 cores, using Intel oneMKL PARDISO package \cite{intelOneMKLPARDISO} with a Python interface \cite{githubGitHubDwfmarchantpyMKL, linear-solver}. Shown in blue is the standard approach as reference where Eq.~\ref{Ax=b} and Eq.~\ref{adjoint_eq} are solved from scratch every iteration by repeating symbolic and numerical factorizations. The green curve demonstrates our improved approach where we compute and cache the numerical factorizations of the forward simulations at different wavelengths and reuse them in the adjoint simulations. Here, the caching of numerical factors reduces the solution of the adjoint system in Eq.~\ref{adjoint_eq} to only a back substitution operation. Finally, the red curve shows optimization performance when both the symbolic and numerical factorizations are cached. Specifically, the additional caching of the symbolic factorization enables its reuse for all subsequent iterations in the optimization process. As a result of these cached factors, while the standard method (blue) reaches our reference objective in \SI{833}{s} of simulation time, the symbolic and numerical factorization-cached approach (red) achieves the same objective in \SI{179}{s}, presenting an improvement of 4.7 times.

The computational advantage is detailed in Fig.~\ref{fig:figure3}(b) with a comparison of the number of factorization and back substitution operations for non-cached and cached optimizations. For the optimization of this device, each iteration runs 1 forward and 1 adjoint simulation at 10 separate target wavelengths, resulting in a total of 20 simulations. Because these factorizations are re-computed in each iteration when no caching is used, these 20 simulations correspond to performing 20 symbolic factorizations, 20 numerical factorizations, and 20 back substitution operations. On the other hand, when factorizations are cached, the symbolic factorization is only computed once during the first iteration. Moreover, even though the numerical factorization is computed anew for every wavelength in the forward simulation of an iteration, these are then cached and reused in the adjoint simulations of the same iteration. Therefore, our caching approach requires only 10 numerical factorizations and 20 back substitutions for all iterations after the first one. As a result, while the standard approach requires \SI{5.09}{s} of computational time for the necessary factorizations and back substitutions at each one of 164 iterations, our factorization caching approach only requires  \SI{1.28}{s} for the first iteration, and \SI{1.12}{s} for all following iterations. This enhancement illustrates the significant computational capabilities enabled by factorization caching in the design of multi-wavelength photonic devices.
\section{Multi-Input and Broadband Spatial Mode Multiplexer Demonstration with Factorization Caching}
\begin{figure}[t]
\centering\includegraphics[width=\linewidth]{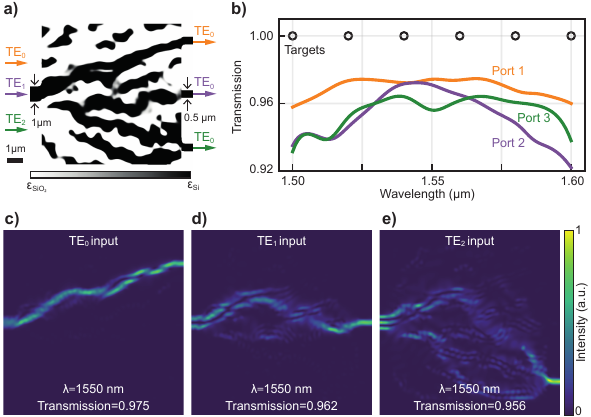}
\caption{\textbf{(a)} Mode multiplexer geometry. \textbf{(b)} Transmission targets (shown with black circles on target wavelengths) and the resulting transmission values for TE\textsubscript{0} mode on the output ports. \textbf{(c-e)} Electric field intensity distribution of mode multiplexer for TE\textsubscript{0}, TE\textsubscript{1}, and TE\textsubscript{2} input modes at \SI{1550}{\nano\metre}.}
\label{fig:figure4}
\end{figure}

The computational scaling achieved by reusing cached factorizations is even more significant in photonic devices that process multiple optical inputs. To demonstrate this, we design a 1 $\times$ 3 broadband silicon photonic mode multiplexer with a \SI{10}{\micro\metre}$\times$\SI{10}{\micro\metre} device footprint as shown in Fig.~\ref{fig:figure4}(a). Here, as the target objective involves multiple modes in addition to multiple wavelengths, higher degrees of design freedom and a larger device footprint arenecessary. In this device, a multi-mode optical input (TE\textsubscript{0}, TE\textsubscript{1}, and TE\textsubscript{2}) is demultiplexed into fundamental modes at three separate outputs, for a broad wavelength range from \SI{1500}{\nano\metre} to \SI{1600}{\nano\metre}. The objective function $F(\mathbf{E})$ is defined as the broadband difference between calculated and target transmissions of fundamental mode outputs at specified ports for each input mode. In each optimization step, forward and adjoint FDFD simulations of TE\textsubscript{0}, TE\textsubscript{1}, and TE\textsubscript{2} inputs are run at 6 target wavelengths. Analogous to the wavelength filter design, a smoothing filter with \SI{200}{\nano\metre} radius and a sigmoid-like projection were performed on the device geometry. The transmission for all three inputs are plotted in Fig.~\ref{fig:figure4}(b), demonstrating an insertion loss of at most 0.35dB (92.2\% transmission) and a broadband mode-conversion operation over the entire specified 1500-\SI{1600}{\nano\metre} spectrum. The corresponding electric fields with TE\textsubscript{0}, TE\textsubscript{1}, and TE\textsubscript{2} input modes are plotted in Fig.~\ref{fig:figure4}(c) through Fig.~\ref{fig:figure4}(e), after an objective function of $2.5 \times 10\textsuperscript{-3}$ is reached in 333 iterations. 

\begin{figure}[t]
\centering\includegraphics[width=\linewidth]{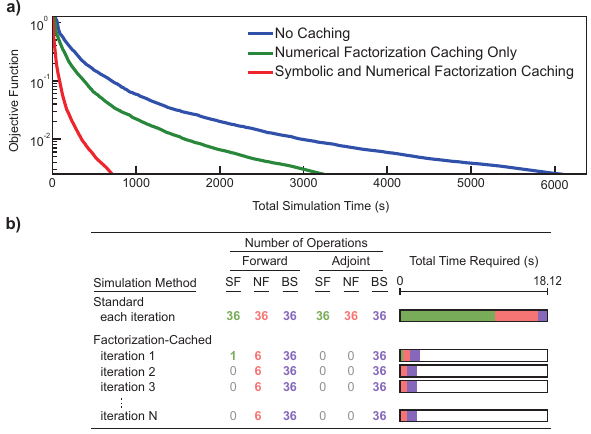}
\caption{\textbf{(a)} Objective function throughout the optimization of a mode multiplexer as a function of total simulation time with different factorization caching approaches. \textbf{(b)}  Numbers of symbolic factorization (SF), numerical factorization (NF) and back substitution (BS) operations in each iteration for no caching and our factorization caching method. Bar plots visualize the total simulation and the corresponding SF, NF and BS runtimes per iteration.
}
\label{fig:figure5}
\end{figure}

We perform a similar analysis on this broadband mode multiplexer by repeating the design process with and without factorization caching, and plot the resulting objective as a function of total simulation time in Fig.~\ref{fig:figure5}(a). The blue, green, and red curves represent the total simulation times for no caching, numerical factorization caching, and both symbolic \& numerical factorization caching approaches, as before. Here, while the standard method (blue) reaches our reference objective in \SI{6076}{s} of simulation time, the same objective is achieved in only \SI{712}{s} when both the symbolic and numerical factorizations are cached (red). This result represents a computational speedup of over 8.5 times over the standard approach, achieving even better scaling than the spectral duplexer we demonstrated above.

The enhanced computational scaling in this device stems from the further advantages enabled by our cached factorizations for the solutions of linear systems with multiple different inputs. These advantages are illustrated by the number of specific operations shown in Fig.~\ref{fig:figure5}(b) for standard and factorization-cached methods. In the standard approach where Eq.~\ref{Ax=b} and Eq.~\ref{adjoint_eq} are re-solved for every iteration at each specified wavelength and input mode, a total of $2\times6\times3=36$ system solutions are required. Performing a total of 36 symbolic factorizations, 36 numerical factorizations, and 36 back substitution operations requires \SI{18.12}{s} of computational time for each iteration of the optimization of this device. In contrast, caching these factorizations for subsequent forward and adjoint simulations allows for the stored symbolic factorization to be reused throughout, and also the stored numerical factorizations to be reused across multiple different inputs at each wavelength. As a result, after the first iteration, only 6 numerical factorizations and 36 back substitutions are required per iteration. Compared to the standard no-caching approach, our method eliminates over 83\% of the numerical factorizations and all subsequent symbolic factorizations, resulting in computational times of \SI{2.40}{s} for the first iteration, and only \SI{2.08}{s} for each subsequent iteration. This clear computational advantage illustrates the benefit of using cached factorizations and how they can significantly enhance the design of multi-input and multi-wavelength optical devices.
\begin{figure*}[t]
\centering\includegraphics[width=0.8\linewidth]{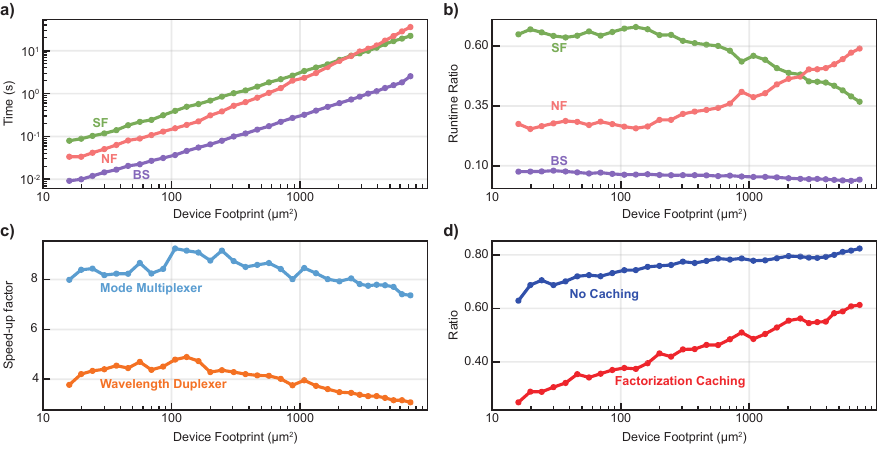}
\caption{\textbf{(a)} Individual runtimes of symbolic factorization (SF), numerical factorization (NF) and back substitution(BS) operations as a function of device footprint. \textbf{(b)} The ratio of individual operation runtimes (SF, NF, BS) to the total simulation time (SF+NF+BS). \textbf{(c)} Speed-up factor for mode multiplexer and wavelength duplexer devices, calculated as the ratio of standard method and factorization-cached method simulation runtimes. \textbf{(d)} The portion of total optimization time used for running FDFD simulations for no-caching and factorization-cached approaches, for the design of the wavelength duplexer.}
\label{fig:figure6}
\end{figure*}

\section{Computational Scaling as a Function of Device Footprint}

As the design footprint directly controls the degrees of freedom of a photonic device and its functional capabilities, it is one of the most critical design parameters in inverse photonic design. Here, we analyze how our resulting computational advantages through factorization caching scale with the device footprint. To do so, we run simulations on devices with footprints ranging from \SI{4}{\micro\metre} $\times$ \SI{4}{\micro\metre} to \SI{85}{\micro\metre} $\times$ \SI{85}{\micro\metre}, with an additional \SI{1}{\micro\metre} on all four sides between the device and the simulation boundaries. Using a \SI{25}{\nano\metre} spatial discretization, these footprints result in system matrices of size from $240^2 \times 240^2$ to $3480^2 \times 3480^2$.  We plot in Fig.~\ref{fig:figure6}(a) the computational time required for three sub-procedures for a single simulation including the symbolic factorization, numerical factorization, and back substitution operations. Fig.~\ref{fig:figure6}(b) shows the ratio of the time required for each one of these operations to their total runtime. From these results, we conclude that while symbolic factorization takes approximately twice as long as numerical factorization with smaller devices (and smaller corresponding system matrices), numerical factorization requires progressively longer times for larger devices. This result reveals that caching the symbolic factorization alone can eliminate over half of the required simulation time, using typical device footprints in inverse photonic design (<\SI{100}{\micro\metre ^2}) \cite{piggott2015inverse, molesky2018inverse, hughes2018adjoint, lalau2013adjoint}. Furthermore, by caching both symbolic and numerical factorizations, it is possible to reduce the necessary computational time to the point where back substitution is the only remaining operation, which corresponds to only 4-8\% of the simulation time, depending on the device footprint. These results indicate that our factorization caching methods can provide significant speedups in simulations and optimizations for a broad range of device footprints and resulting system matrices.

The total enhancement in simulation speedup is plotted in Fig.~\ref{fig:figure6}(c) for the two devices demonstrated, as a function of the device footprint. The simulations for wavelength duplexer with factorization caching are over 4.7 times faster than standard no-caching approach for devices smaller than  \SI{10}{\micro\metre} $\times$ \SI{10}{\micro\metre}, and are about 3.1 times faster for much larger devices (up to \SI{85}{\micro\metre} $\times$ \SI{85}{\micro\metre}). Even though the symbolic factorization is completely eliminated after the first iteration, the computational enhancement decreases slightly with increasing device size, as the necessary numerical factorizations in forward and adjoint simulations start to occupy the majority of total iteration times with larger system matrices. A similar trend is observed for the broadband mode multiplexer where the speed-up factor of 9.2 for the typical device footprint reduces to 7.3 for the largest device we tested (\SI{85}{\micro\metre} $\times$ \SI{85}{\micro\metre}). While the factorization caching approach may not achieve the exact same level of computational enhancement for these much larger footprints as it does for standard ones, Fig.~\ref{fig:figure6}(c) demonstrates that our approach still leads to a significant reduction in simulation time regardless of footprint in the scope of the inverse design applications. Lastly, in Fig.~\ref{fig:figure6}(d), we present a comparison of the simulation time to the total time required for each iteration, which also includes operations such as smoothing and binarization of the device geometry, gradient computations, and the optimization algorithm itself. As plotted by the blue curve, over 62\% of the total iteration time is spent on the forward and adjoint physical simulations themselves, when no caching is employed. This ratio reaches as high as 82\% with larger devices with footprints of \SI{85}{\micro\metre} $\times$ \SI{85}{\micro\metre}. With our factorization caching approach (red curve), only 24\% of the time required for each iteration is used for these simulations in \SI{4}{\micro\metre} $\times$ \SI{4}{\micro\metre} devices, reaching up to 61\% in \SI{85}{\micro\metre} $\times$ \SI{85}{\micro\metre} devices. As expected, the time required for physical simulations already comprises the vast majority of computational requirements for these much larger devices. This explains why our achieved speedup factors experience a slight decrease with increasing device footprint. These results demonstrate that with factorization caching, linear system solutions can now be completed quicker than the other operations necessary for nanophotonic device design (including image smoothing, gradient calculation, and optimization). Further enhancements through GPU-accelerated image operations or advancements in rapid optimization algorithms can further accelerate inverse design frameworks.

\section{Conclusion}

In summary, we have demonstrated a multi-faceted factorization caching approach for computational enhancement of inverse photonic design by leveraging the symbolic and numerical structure of the system matrix. Our results highlight the effectiveness of factorization caching in reducing the computational cost associated with device simulation and optimization processes. As we introduce no changes or approximations in the underlying system of equations, our method remains generalizable to a large class of photonic design problems with the system solutions remaining exact, physically accurate, and data-independent. Coupled with state-of-the-art optimization algorithms, this approach can enable the design of novel, large-scale, and multi-purpose nanophotonic devices in a variety of integrated photonics platforms. Finally, our factorization caching method can even extend beyond photonics, as it can be applied to many other computational applications requiring repetitive solutions of structurally similar system matrices.

\begin{acknowledgments}
This work is supported by Scientific and Technological Research Council of Turkey (TUBITAK) under grant number 119E195.
\end{acknowledgments}

\section*{DATA AVAILABILITY}
The data that support the findings within this manuscript are available from the corresponding author upon reasonable request.

%\section*{CODE AVAILABILITY}
%The code that supports the findings within this manuscript are available in the cited references.

\section*{COMPETING INTEREST}
The authors declare no competing interests.

%% took the line below from internet to remove the black horizontal line 
\def\bibsection{\section*{References}}
\bibliography{main}% Produces the bibliography via BibTeX.

\end{document}